\documentstyle[epsfig]{aprim}

\newif\ifAMStwofonts

\ifoldfss
  \ifCUPmtlplainloaded \else
    \NewTextAlphabet{textbfit} {cmbxti10} {}
    \NewTextAlphabet{textbfss} {cmssbx10} {}
    \NewMathAlphabet{mathbfit} {cmbxti10} {} 
    \NewMathAlphabet{mathbfss} {cmssbx10} {} 
  \fi
  \ifAMStwofonts
    \ifCUPmtlplainloaded \else
      \NewSymbolFont{upmath} {eurm10}
      \NewSymbolFont{AMSa} {msam10}
      \NewMathSymbol{\upi}     {0}{upmath}{19}
      \NewMathSymbol{\umu}     {0}{upmath}{16}
      \NewMathSymbol{\upartial}{0}{upmath}{40}
      \NewMathSymbol{\leqslant}{3}{AMSa}{36}
      \NewMathSymbol{\geqslant}{3}{AMSa}{3E}

    \fi
  \fi
\fi 

\ifnfssone
  \newmathalphabet{\mathit}
  \addtoversion{normal}{\mathit}{cmr}{m}{it}
  \addtoversion{bold}{\mathit}{cmr}{bx}{it}
  \newmathalphabet{\mathbfit} 
  \addtoversion{normal}{\mathbfit}{cmr}{bx}{it}
  \addtoversion{bold}{\mathbfit}{cmr}{bx}{it}
  \newmathalphabet{\mathbfss} 
  \addtoversion{normal}{\mathbfss}{cmss}{bx}{n}
  \addtoversion{bold}{\mathbfss}{cmss}{bx}{n}
  \ifAMStwofonts
    \ifCUPmtlplainloaded \else
      \UseAMStwoboldmath
      \makeatletter
      \new@mathgroup\upmath@group
      \define@mathgroup\mv@normal\upmath@group{eur}{m}{n}
      \define@mathgroup\mv@bold\upmath@group{eur}{b}{n}
      \edef\UPM{\hexnumber\upmath@group}
      \new@mathgroup\amsa@group
      \define@mathgroup\mv@normal\amsa@group{msa}{m}{n}
      \define@mathgroup\mv@bold\amsa@group{msa}{m}{n}
      \edef\AMSa{\hexnumber\amsa@group}
      \makeatother
      \mathchardef\upi="0\UPM19
      \mathchardef\umu="0\UPM16
      \mathchardef\upartial="0\UPM40
      \mathchardef\leqslant="3\AMSa36
      \mathchardef\geqslant="3\AMSa3E
    \fi
  \fi
\fi 

\ifnfsstwo
  \DeclareMathAlphabet{\mathbfit}{OT1}{cmr}{bx}{it}
  \SetMathAlphabet\mathbfit{bold}{OT1}{cmr}{bx}{it}
  \DeclareMathAlphabet{\mathbfss}{OT1}{cmss}{bx}{n}
  \SetMathAlphabet\mathbfss{bold}{OT1}{cmss}{bx}{n}
  \ifAMStwofonts
    \ifCUPmtlplainloaded \else
      \DeclareSymbolFont{UPM}{U}{eur}{m}{n}
      \SetSymbolFont{UPM}{bold}{U}{eur}{b}{n}
      \DeclareSymbolFont{AMSa}{U}{msa}{m}{n}
      \DeclareMathSymbol{\upi}{0}{UPM}{"19}
      \DeclareMathSymbol{\umu}{0}{UPM}{"16}
      \DeclareMathSymbol{\upartial}{0}{UPM}{"40}
      \DeclareMathSymbol{\leqslant}{3}{AMSa}{"36}
      \DeclareMathSymbol{\geqslant}{3}{AMSa}{"3E}
    \fi
  \fi
\fi 

\ifCUPmtlplainloaded \else
  \ifAMStwofonts \else 
    \def\upi{\pi}
    \def\umu{\mu}
    \def\upartial{\partial}
  \fi
\fi

\title[Thermal Instability and Fluctuations]{Linear Thermal Instability
and Fluctuations in Molecular Clouds}

\author[Nejad-Asghar et al.]
       {M. Nejad-Asghar$^1$ and J. Ghanbari$^2$\\
        $^1$Department of Physics, Damghan University of Basic Sciences,
Damghan, Iran\\
        $^2$Department of Physics, Ferdowsi University of Mashhad,
Mashhad, Iran}
\date{}

\pagerange{\pageref{firstpage}--\pageref{lastpage}}
\pubyear{2005}

\begin{document}

\maketitle

\label{firstpage}

\begin{abstract}
Evidence of small-scale condensations in the magnetic molecular
clouds has been accumulating over the past decades through radio
and optical/ultraviolet observations. The origin and shape of
these small-scale condensations is a disputable issue.
Nejad-Asghar \& Ghanbari (2004 hereafter NG) have recently studied
the effect of the linear thermal instability on the formation of
fluctuations in molecular clouds. The authors inferred that under
certain conditions (e.g., depending on expansion or contraction
of the background) thermal instability and ambipolar diffusion
can produce spherical, oblate, or prolate condensations.
\end{abstract}


\section{Introduction}

Firstly, about presence of these fluctuations. We have two method
to find them: (1) direct imaging of nearby clouds reveals
substructures on scales to lengths of $0.01pc$ and masses of
$0.01M_\odot$ \cite{peng98,saka03}, (2) studies of the time
variability of absorption lines indicates the presence of
fluctuations on scales of $10^{-4} pc$ (5-50 AU) and masses of
$10^{-9} M_\odot$ \cite{moor95,bois05}.

How these fluctuations may be formed and is the thermal
instability important? To answer the question, first, we must
consider time-scales. According to the prior results of Gilden
\shortcite{gild84}, thermal instability time-scale in molecular
clouds is in the order of $10^3-10^4$ year. On the other hand,
Larson \shortcite{lars81} showed that dynamical time-scale for
transient structure of molecular clouds is in order of $10^7$
year. Thermal instability time-scale is very smaller than
dynamical time-scale. Thus, it is reasonable to consider thermal
instability as an important mechanism in formation of small
fluctuations in molecular clouds.

It is now accepted that the magnetic field is very important in
dynamical evolution of all ISM. It will affect on ions directly
and on neutral particles indirectly, via collision with ions. In
molecular clouds, ionization degree is very small, thus, drift of
neutral particles between tied ions is important. It can heat the
medium. In this way, for complete investigation of the thermal
instability in molecular clouds, we must consider plasma drift
(ambipolar diffusion).

\section{Net Cooling Function and Perturbation}

The most important parameter in thermal instability is the net
cooling function. A good fitted result for cooling in molecular
clouds is $\Lambda(\rho,T)=\Lambda_0 \rho^\delta T^\beta$
\cite{gold78,neuf95}, where the values of $\delta$ and $\beta$ is
given by the Fig.~1 of NG.

About heating rate, we turn our attention to gravitational
expantion/contraction work and ambipolar diffusion heating rate.
The gravitational heating rate is found by setting the rate of
compressional/expansional work per particle, equal to the rate of
change of gravitational energy per particle. For ambipolar
diffusion heating rate, we have
\begin{equation}
\Gamma_{AD}=\frac{\textbf{\textit{f}}_d\cdot\textbf{\textit{v}}_d}{\rho_n}
\end{equation}
where $\textbf{\textit{f}}_d=\eta\epsilon \rho^{1+\nu}$ is the
drag force (per unit volume) with $\eta$ as collision drag, and
$\textbf{v}_d$ is the drift velocity of ions relative to neural
particles. In order of magnitude, if $\kappa B_0$ changes on a
typical scale of $\lambda$, and if we choose $\rho_i \sim
\epsilon \rho_n^\nu$ \cite{umeb80}, ambipolar diffusion heating
rate, in a good estimate, is given by
\begin{equation}
\Gamma_{AD}=\Gamma_0''\rho^{-(2+\nu)};\quad
\Gamma_0''\equiv\frac{(\kappa
B_0)^4}{16\pi^2\eta\epsilon\lambda^2}.
\end{equation}
In this way, the net cooling function in molecular clouds is
given by (for more detail see NG)
\begin{equation}
  \Omega(\rho,T)=\Lambda_0 \rho^\delta T^\beta
  - \Gamma_0'\rho^{3/2}- \Gamma_0''\rho^{-(2+\nu)}.
\end{equation}
Instability condition depends on the different values of $\beta$,
$\delta$, and $\nu$.

We use the perturbation method to investigate the occurrence of
thermal instability in molecular clouds and formation of
fluctuations. We choose an expanding/contracting molecular cloud
with uniform magnetic field $\textbf{B}_0$. Perturbation on this
medium results to a five-degree linear characteristic equation,
which its coefficients depend on the importance of ambipolar
diffusion and gravitational heating rate
\begin{equation}
\xi\equiv\frac{\Gamma_{AD}}{\Lambda(n_0,T_0)},\quad
\chi\equiv\frac{\Gamma_{grav}}{\Lambda(n_0,T_0)},
\end{equation}
and some time-scales that are investigated in details at
subsection 3.2 of NG. By introducing the non-dimensional
quantities
\begin{equation}
y\equiv h\tau_s,\; \sigma_\rho\equiv\frac{\tau_s}{\tau_{cT}},\;
\sigma_T\equiv\frac{\tau_s}{\tau_{c\rho}}
+\frac{\tau_s}{\tau_K},
\end{equation}
\begin{equation}
\alpha\equiv(\frac{\tau_s}{\tau_{AL}})^2,\;
D\equiv\frac{\tau_s}{\kappa^2\tau_{AD}},\;
G_g\equiv(\frac{\tau_s}{\tau_g})^2,\;
E_e\equiv\frac{\tau_s}{\tau_e},
\end{equation}
we use the Laguerre method to find five roots of the
characteristic equation \cite{neja05}.

\section{Results and Prospects}

If we consider the effect of the self gravity without
expansion/contraction of the background, the isobaric instability
criterion (line $OA$ of Fig.~1) is modified, because, self-gravity
causes to increase the internal pressure.

\input{epsf}
\begin{figure} 
 \centerline{{\epsfxsize=4cm\epsffile{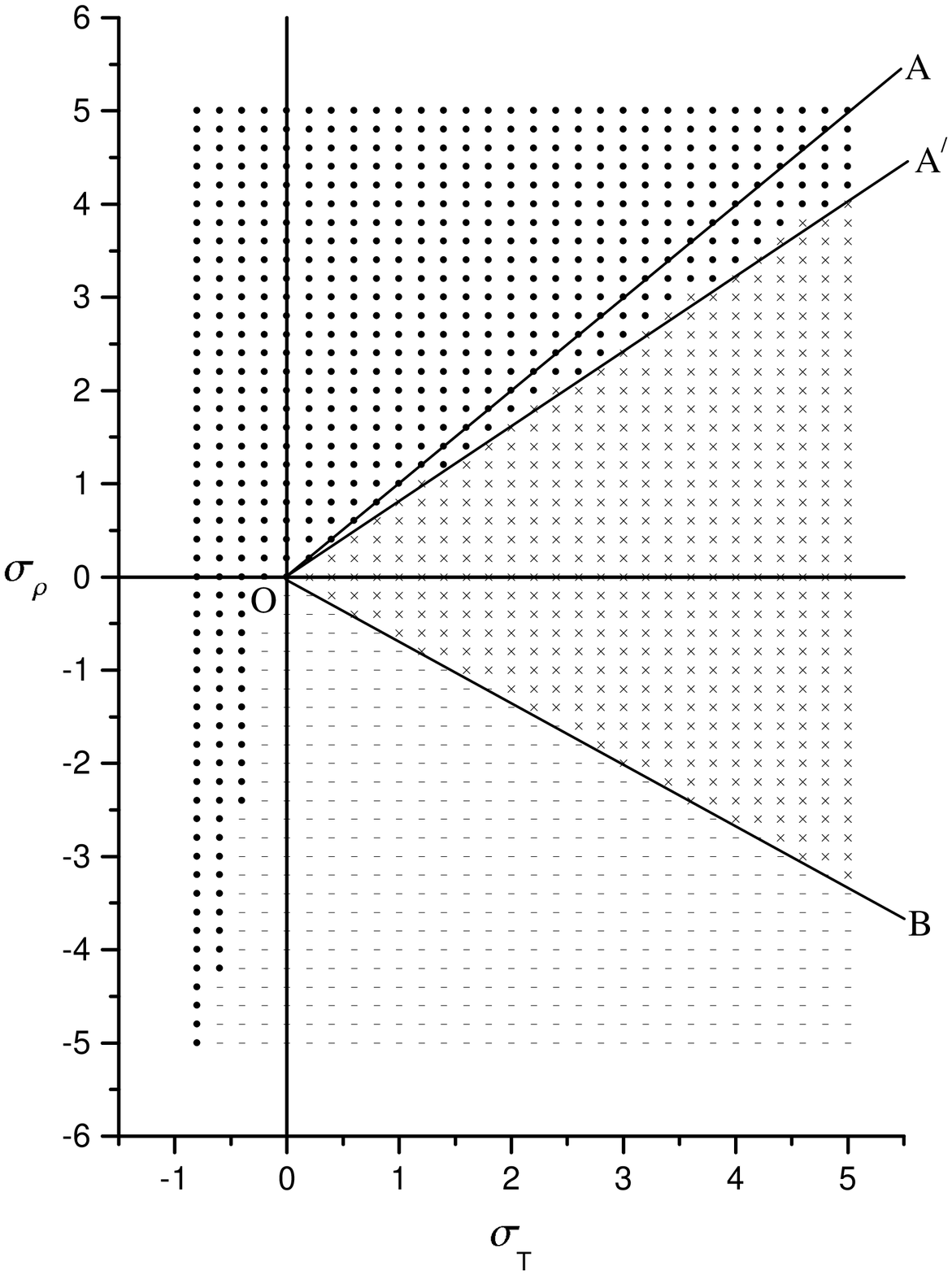}}}
 \caption[]{Regions of stability($\times$),
 spherical instability($\bullet$), and oblate instability($-$)
 for the case of $G_g=0.1$ and $E_e=0$, for a typical value of
 $\alpha=5.0$ and $D=1.0$.}
\end{figure}

If the background is expanding ($E_e>0$), its expansion energy
causes to stabilize the medium. This case is shown in Fig.~2 for
a typical value of $\alpha=5.0$ and $D=1.0$. In the isentropic
instability criterion (line $OB$), expansion energy causes to
stabilize the medium in the direction of the magnetic field and
perpendicular to it. In the isobaric instability criterion (line
$OA$), it only causes to stabilize the medium in perpendicular to
the magnetic field, corresponding to decreased pressure via
ion-neutral friction.

\input{epsf}
\begin{figure} 
 \centerline{{\epsfxsize=3.8cm\epsffile{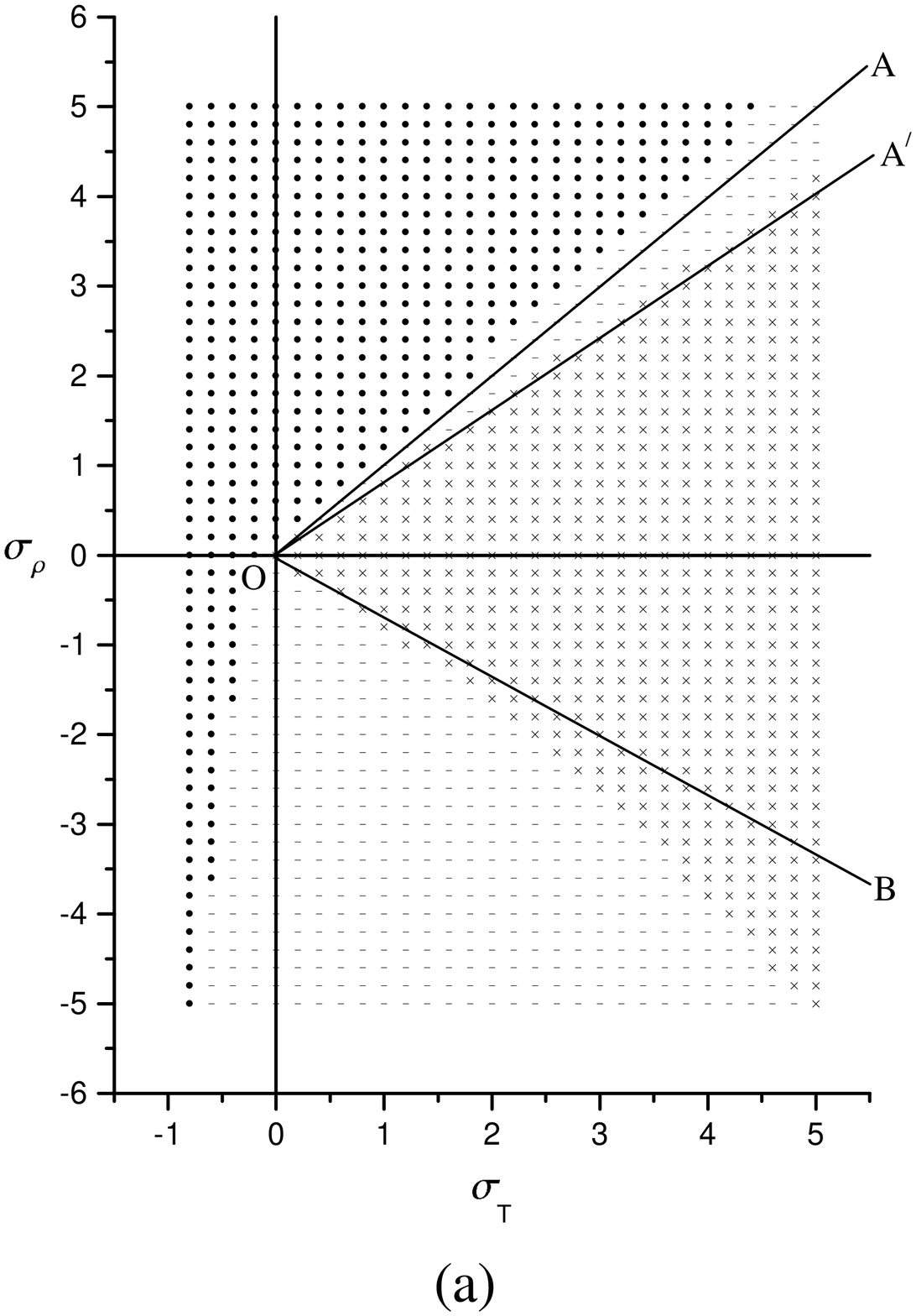}}
   {\epsfxsize=3.8cm\epsffile{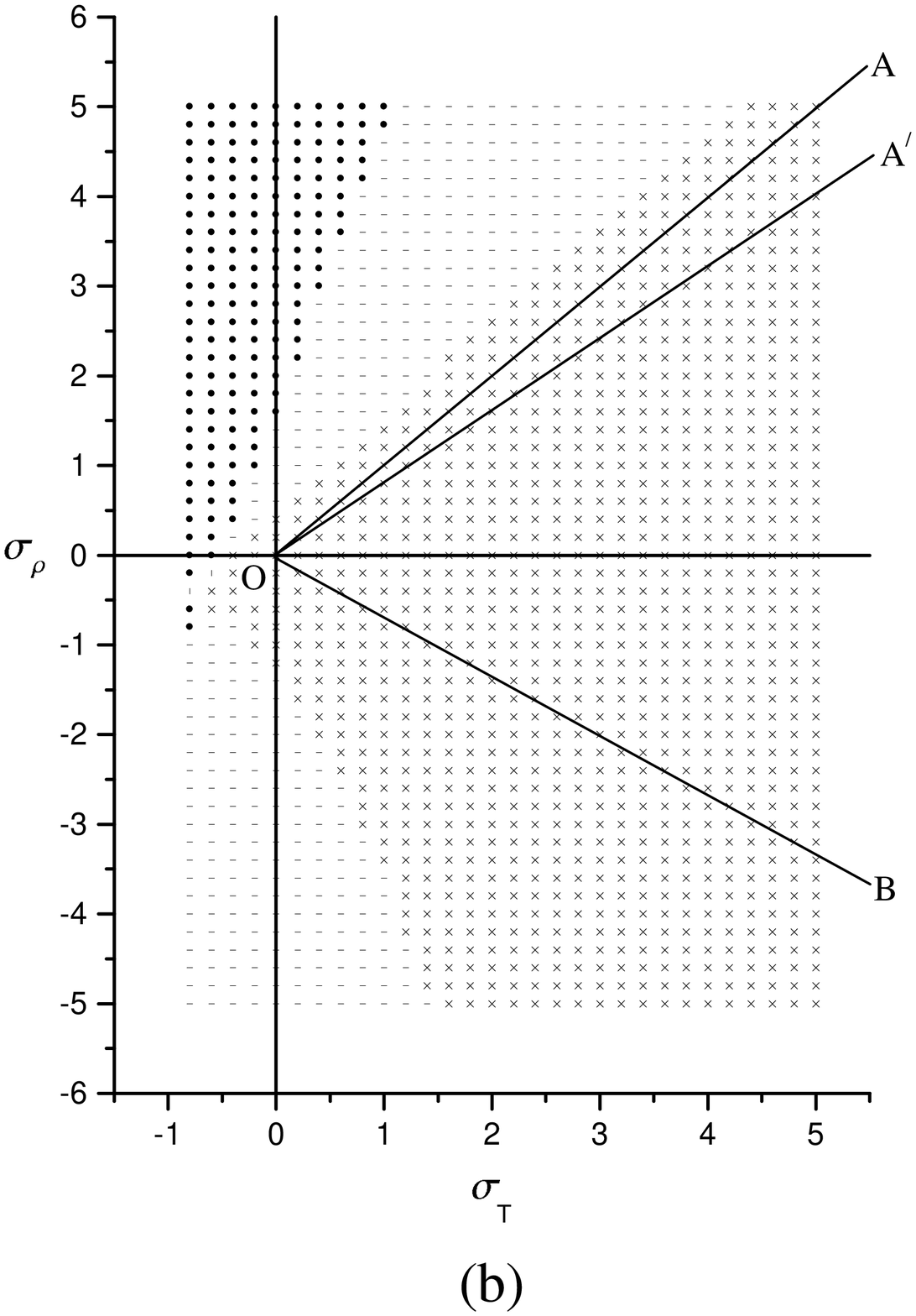}}}
 \caption[]{Regions of stability( $\times$),
 spherical instability($\bullet$), and oblate instability($-$)
 in the expanding background, for a typical value of
 $\alpha=5.0$ and $D=1.0$, with (a) $E_e=0.01$ and $G_g=0.1$, and
 (b) $E_e=0.1$ and $G_g=0$.}
\end{figure}

For contracting background ($E_e<0$), contraction energy injected
to the medium, thus, its stability is decreased and converted to
a prolate instability. Diffusion of neutrals relative to the
freezed ions in the perpendicular direction of the magnetic field
is the reason of this prolate instability. This case is shown in
Fig.~3 for a typical value of $\alpha=5.0$ and $D=1.0$. When the
parameters of a magnetic molecular cloud set, locally, in this
region of $\sigma_T-\sigma_\rho$ plane, prolate condensation may
be produced via thermal instability.

\begin{figure} 
 \centerline{{\epsfxsize=3.8cm\epsffile{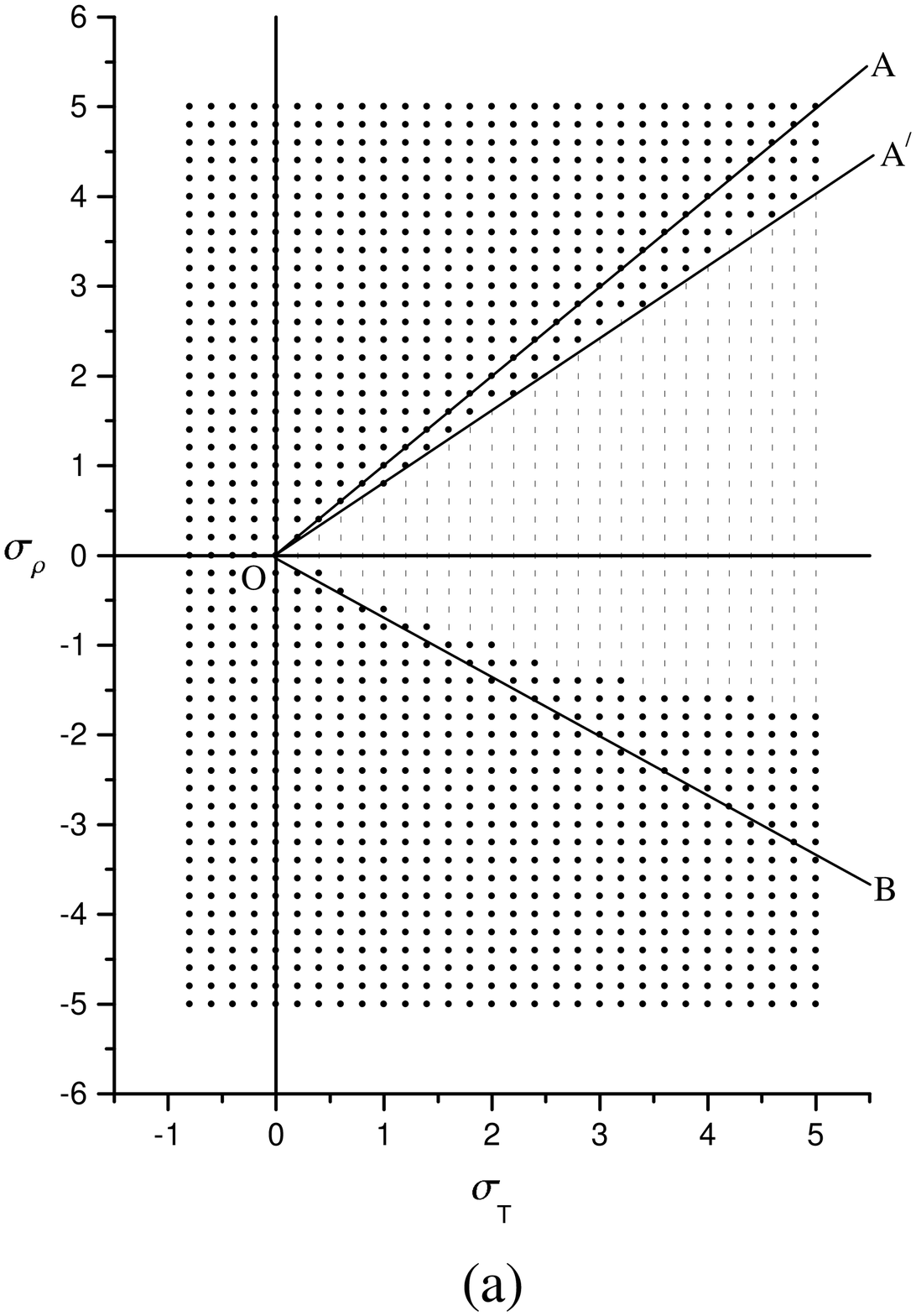}}
   {\epsfxsize=3.8cm\epsffile{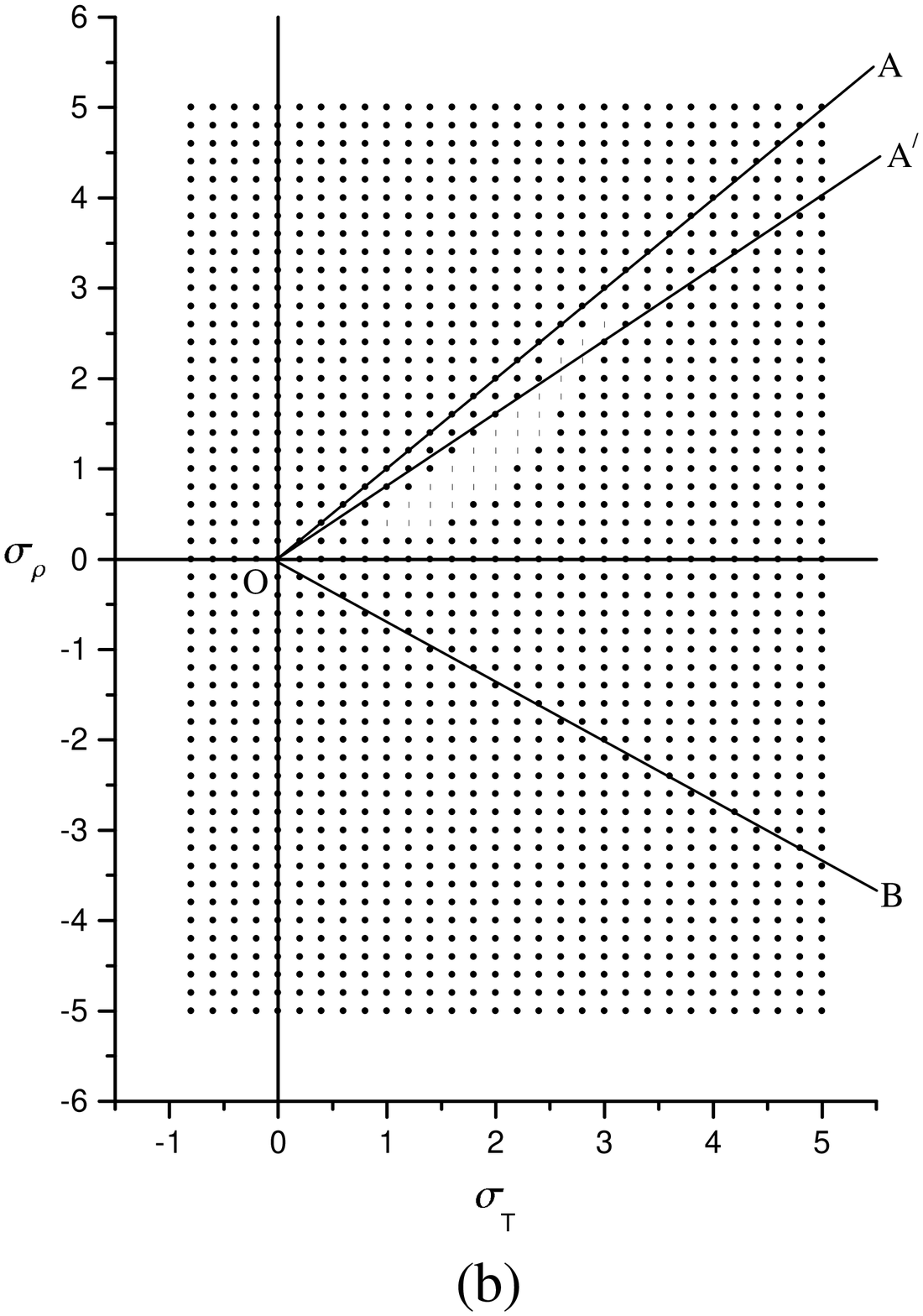}}}
 \caption[]{Regions of spherical instability($\bullet$), and prolate
 instability($|$) in the contracting background, for a typical value of
 $\alpha=5.0$ and $D=1.0$, with (a) $E_e=-0.01$ and $G_g=0.1$, and
 (b) $E_e=-0.1$ and $G_g=0$.}
\end{figure}

Investigation of this problem in the linear prospect is not
completely OK, because, ambipolar diffusion is a nonlinear effect
and density fluctuation ratios are in the order of 10. But, we
see that linear study can give us a good idea about the
conditions for occurrence of thermal instability. For complete
study we must work in nonlinear regime.

\label{lastpage}

\clearpage

\end{document}